\documentclass{aa}
\usepackage{graphicx}
\usepackage{txfonts}
\usepackage{color}

\newcommand{\mbh}{M_{\bullet}}
\newcommand{\msun}{M_{\sun}}
\newcommand{\mast}{M_{\star}}
\newcommand{\md}{M_{\mathrm{d}}}
\newcommand{\rg}{R_{\mathrm{g}}}
\newcommand{\rh}{R_{\mathrm{h}}}
\newcommand{\rd}{R_{\mathrm{d}}}
\newcommand{\rt}{R_{\mathrm{t}}}
\newcommand{\rp}{R_{\mathrm{p}}}
\newcommand{\rast}{R_{\star}}
\newcommand{\rsun}{R_{\sun}}
\newcommand{\rmin}{R_{\rm{}min}}
\newcommand{\bs}{\!}
\newcommand{\der}{\mathrm{d}}
\newcommand{\mbf}[1]{\mbox{\boldmath$#1$}}

\begin{document}
\title{Enhanced activity of massive black holes by stellar capture assisted by a self-gravitating accretion disc}
\author{V.~Karas\inst{1} \and L.~\v{S}ubr\inst{\,2,3}}
\institute{$^1$~Astronomical Institute, Academy of Sciences, Bo\v{c}n\'{\i}~II 1401, CZ-14131~Prague, Czech Republic\\
$^2$~Argelander Institut f\"ur Astronomie, University of Bonn, Auf dem H\"ugel~71, D-53121~Bonn, Germany\\
$^3$~Astronomical Institute, Charles University, Faculty of Mathematics and Physics, V~Hole\v{s}ovi\v{c}k\'ach~2, CZ-18000~Prague, Czech Republic}
\authorrunning{V.~Karas \& L.~\v{S}ubr}
\titlerunning{Enhanced activity of massive black holes by stellar capture\ldots}
\date{Received 19 July 2006; accepted 15 April 2007}
\abstract{}%
{We study the probability of close encounters between stars from a nuclear
cluster and a massive black hole ($10^4\msun\lesssim\mbh\lesssim10^8\msun$). The 
gravitational field of the system is dominated by the black hole
in its sphere of influence. It is further modified by the cluster mean 
field (a spherical term) and a gaseous disc/torus 
(an axially symmetric term)
causing a secular evolution of stellar orbits via Kozai oscillations.
Intermittent phases of large eccentricity increase the chance that
stars become damaged inside the tidal radius of the central hole. Such
events can produce debris and lead to recurring episodes of enhanced 
accretion activity.}
{We introduce an effective loss cone and
associate it with tidal disruptions during the high-eccentricity phases
of the Kozai cycle. By numerical integration of the trajectories forming
the boundary of the loss cone we determine its shape and volume. We also 
include the effect of relativistic advance of pericentre.}
{The potential of the disc has the efffect of enlarging the loss cone
and, therefore, the predicted number of tidally disrupted stars should grow by factor 
of $\simeq10^2$. On the other hand, the effect of the cluster mean potential
together with the relativistic pericentre advance act against the eccentricity 
oscillations. In the end we expect the tidal disruption events to be approximately
ten times more frequent in comparison with the model in which the three
effects -- the cluster mean field, the relativistic pericentre advance, and
the Kozai mechanism -- are all ignored. The competition of different influences
suppresses the predicted star disruption rate as the black hole mass increases.
Hence, the process under consideration is more important for intermediate-mass
black holes, $\mbh\simeq10^4\msun$.}{}
\maketitle

\section{Introduction}
According to the present-day consensus, the observational evidence shows
black holes populating the mass spectrum in two distinct intervals:
stellar-mass black holes, $3\msun\lesssim\mbh\lesssim20\msun$, that
originate from collapse of massive stars and are revealed as members of
binary systems; and supermassive black holes (SMBHs;
$10^6\msun\lesssim\mbh\lesssim10^9\msun$) hosted in nuclei of many
galaxies including our own. The existence of intermediate-mass black
holes (IMBHs; $10^2\msun\lesssim\mbh\lesssim10^5\msun$) is an open
issue, likewise their observational consequences and the form of the
mass spectrum that might be filling the gap between the two
well-established categories of black holes.

Over the last four decades, a picture of galactic nuclei has emerged in
which black holes are embedded in a dense stellar system and their
masses are increasing by accretion of gas from an accretion disc,
gradually inflowing towards the horizon (Begelman \& Rees \cite{begelman78};
Volonteri \& Rees \cite{volonteri05}). Remnants of stars tidally
disrupted near a SMBH also contribute as a source of material (Zhao et
al.\ \cite{zhao02}). Previously we demonstrated (Vokrouhlick\'y \& Karas
\cite{vokrouhlicky98}; \v{S}ubr et al.\ \cite{subr04})
that stars of the nuclear cluster can undergo episodes of very large
orbital eccentricity if they interact with a self-gravitating disc near
a SMBH. This mechanism sets stars on very elongated trajectories with
small periapses, although the overall tendency of the gas-assisted drag
acts in the opposite way -- towards the orbital circularization (Syer
et al.\ \cite{syer91}; Vokrouhlick\'y \& Karas \cite{vokrouhlicky93}).

In this paper we study the fraction of stars that are set to highly
eccentric  orbits as a function of the black hole mass. We suggest this
problem is relevant in the context of coexistence of a massive  black
hole with a surrounding cluster of stars. Exactly because of large
elongation of the orbits, the fate of the remnant gas from the
tidally disrupted stars is very uncertain
and we do not attempt to solve this problem in its entirety. Instead, we
concentrate our attention on a single aspect -- the mutual gravitational
interaction of stars and the gaseous disc. Our model is axially
symmetric. We study how Kozai's phenomenon modifies and enlarges the
black-hole loss cone and how this change depends on $\mbh$. We do take
several subtleties of this scenario into consideration, namely the 
damping effect of the star cluster and the relativistic pericentre 
advance of stellar orbits near the central black hole. These effects
pose a potential threat to our mechanism.

We find that a self-gravitating disc is capable of pushing more stars
towards less massive (i.e.\ less-than-supermassive) black holes.
Therefore, the effect should be particularly relevant for
intermediate-mass black holes, provided they exist embedded within dense
stellar systems and accrete from a gaseous disc or a torus (Madau \&
Rees \cite{madau01}; Miller \& Hamilton \cite{miller02}).

There is a secondary motivation for our investigation: because the
efficiency of the mechanism discussed herein decreases with the black
hole mass increasing (it seems to be overwhelmed entirely by damping
effects above $\mbh\sim10^7\msun$), the present model may be
particularly relevant for an ongoing debate about the origin of
ultra-luminous X-ray sources (ULXs) and their putative connection with
IMBHs (for reviews, see van der Marel \cite{marel04}; Fabbiano
\cite{fabbiano06}). 

ULXs are extra-nuclear point-like X-ray sources with isotropic
luminosities exceeding $10^{39}\,\mbox{erg\,s}^{-1}$. It has been
proposed (Colbert \& Mushotzky \cite{colbert99}; Makishima et al.\
\cite{makishima00}; Matsumoto et al.\ \cite{matsumoto01}; Portegies
Zwart et al.\ \cite{zwart04}; Hopman et al.\ \cite{hopman04}; Baumgardt
et al.\ \cite{baumgardt06}) that some
of them may harbour accreting IMBHs. Within the intermediate black-hole
mass range the proposed mechanism is relatively efficient, and we can
thus speculate that these ULXs might be triggered when a star is damaged
near the tidal radius. This concept seems to be in line also with the
idea that accreting pre-galactic seed holes may could indeed be
detectable as  ULXs (Madau \& Rees \cite{madau01}). Also, it appears to
be in accord with the manifestation of a multi-colour disc component, as
reported in spectra of a sample of ULXs (Miller et al.\ \cite{miller04};
\cite{miller06}; Fabian et al.\ \cite{fabian04}), and it also agrees
with indications  that ULXs are transient (Miniutti et al.\
\cite{miniutti06}), possibly recurring phenomenon and that they are
switched on during phases of active accretion (see Krolik
\cite{krolik04}). However, the evidence for IMBHs is only
circumstantial -- numerous uncertainties persist in the mass estimates
(see the discussion in King et al.\ \cite{king01}; Gon\c{c}alves \&
Soria \cite{goncalves06}), and so our scheme is a mere speculation at
this stage. 

\section{Model and method}
\label{sec:model}
\subsection{Basic equations}
We assume that some kind of an accretion disc surrounds the central 
black hole and defines the equatorial plane of the system. The presence
of the black hole provides a natural length-scale which we will use
hereafter: $\rg\equiv{G\mbh}c^{-2}\simeq1.5\times10^9M_4$~cm,
$M_4\,\equiv\,\mbh/(10^4\msun)$. The disc is taken as axially  symmetric
and relatively light with respect to the central black hole
($\mu\equiv\md/\mbh\ll1$). Nevertheless, it is important that $\mu$ is
greater than zero. To be specific, we consider two examples: (i)~a
constant surface density disc, characterized by its  mass $\md$ and the
outer radius $\rd$, and (ii)~a limiting case of an infinitesimally
narrow ring of mass $\md$ and radius $\rd$. The two cases serve as a
useful test-bed because analytical expressions can be derived for the
perturbation potential, $V_\mathrm{d}(R,z)$, and its first derivatives
(Lass \& Blitzer \cite{lb83}; Hur\'e \& Pierens \cite{hure05}). Notice
also that the massive ring surrounding the central black hole can
represent a time-averaged system with a secondary black hole, sometime
invoked in the framework of the hierarchical SMBH formation in centres
of merging protogalaxies (Volonteri et al.\ \cite{volonteri03};
G\"ultekin et al.\ \cite{gultekin04}). Recently, Ivanov et al.\
(\cite{ivanov05}) applied the averaging technique to study tidal
disruptions of stars in a galaxy centre containing a supermassive binary
black hole. However, these authors neglected relativistic effects, which
exhibit growing importance near the black hole.

The central cluster is bound gravitationally to the black hole and
characterized by distribution function on the space of osculating
elements, $D_\mathrm{f}(a,C_1,e,\omega)$, where $C_1=\eta\cos i\,$ is a
normalized $z$-component of the angular momentum,
$\eta\,\equiv\,(1-e^2)^{1/2}$; $e$ is eccentricity, $a$ is semimajor
axis and $\omega$ is argument of pericentre. The orbit inclination $i$
is  measured with respect to the disc plane. We consider distribution of 
the form
\begin{equation}
 D_{\mathrm{f}}(a,C_1,e,\omega)=Ka^{1/4}e\eta^{-1}
\label{eq:df}
\end{equation}
with $K$ being a normalization constant. Eq.~(\ref{eq:df}) represents
the Bahcall \& Wolf (\cite{bahcall76}) distribution with a random
orientation in inclinations and linear distribution in eccentricities.
The stellar cluster introduces a spherically symmetric perturbation to
the central potential, $V_\mathrm{c}(r)\propto r^{1/4}$.

Length-scales of the model can be related to the mass according to the
empirical relation $\mbh(\sigma)$ (Tremaine et al.~\cite{Tremaine02})
which we extrapolate in a naive way from galactic nuclei down to IMBHs
range: $M_4\simeq10^4\sigma_2^4$, where
$\sigma_2\,\equiv\,\sigma/(200{\rm{}km\,s}^{-1})$ is the velocity
dispersion. This seems to be
justified by kinematical studies of some globular clusters (e.g., M15, Gerssen
et al.\ \cite{gerssen02}; G1, Gebhardt et al.\ \cite{gebhardt05};
47~Tuc, McLaughlin et al.\ \cite{mclaughlin06}) and it enables us to introduce the cusp
radius,
\begin{equation}
\rh\equiv G\mbh\sigma^{-2} = 2.25\times10^8 M_4^{-1/2}\! \rg
= 0.11 M_4^{1/2}\;\mathrm{pc}\,.
\label{eq:Rh}
\end{equation}
We further assume the mass of the cluster $M_\mathrm{c}(\rh)\simeq\mbh$.
Finally, we set the disc/ring radius to be equal to the characteristic
radius of the cluster: $\rd=\rh$. Thenceforth only $\md$ is a free parameter.

The axial symmetry of the problem does not ensure conservation of the
angular momentum; instead, only one of its components is conserved,
so that the orbital eccentricity and inclination of the cluster stars can
secularly evolve and fluctuate. In our model we imagine that
Kozai's phenomenon\footnote{Originally (Kozai \cite{kozai62}; Lidov
\cite{lidov62}), the averaging method was applied 
to study the long-term evolution in the context of the restricted
three-body problem. Naturally, the approach can be readily applied to
discuss the motion in an axially symmetric potential.} is responsible for
oscillations of the orbital elements on the time-scale of
\begin{equation}
T_\mathrm{K} \equiv\frac{\lambda}{2\pi\mu} \left(\frac{\rd}{a}\right)^3\! P 
\simeq 1.6\times 10^3 M_4\, \frac{\lambda}{\mu} \left( \frac{\rd}{a} \right)^3 
\left(\frac{a}{10^8\rg} \right)^{3/2}\!\!\mathrm{yr}.
\label{eq:time_kozai}
\end{equation}
The oscillation period exceeds the orbital period $P(a)$ of each
individual star, $T_\mathrm{K}\gg P=2\pi\,a^{3/2}/(G\mbh)^{1/2}$,
by at least one order of magnitude. Factor $\lambda\simeq1$
includes additional corrections arising from unaccounted effects.

\begin{figure}
\includegraphics[width=\columnwidth]{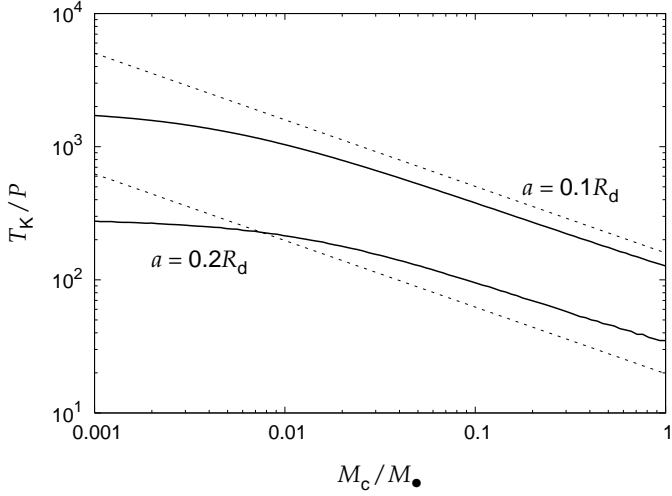}
\caption{Kozai period 
$T_\mathrm{K}$ (solid lines) as a function of mass distributed in the spherical
potential of the nuclear cluster. Time is scaled with the corresponding orbital 
period $P$. Principal parameters of the system are the black hole mass, $M_4=10$, 
and the mass of the ring, $M_\mathrm{d}=0.1\mbh$.
Trajectories were characterised by $C_1=10^{-3}$ and integrated
numerically. Two different values of semi-major axis are given by each
curve. Also plotted is the approximation to $T_\mathrm{K}$ with the
value of $\lambda$ from eq.~(\ref{eq:time_kozai_corrected}), producing
the slope $-1/2$ (dotted lines).}
\label{fig:T_K}
\end{figure}

Precession due to an extended spherically symmetric
potential component accelerates the oscillations. For 
$M_\mathrm{c} \gtrsim 0.1\mbh$ we find the correction 
term to the cycles of eccentricity oscillations,
\begin{equation}
\lambda \simeq 0.1 \left( \frac{M_\mathrm{c}}{\mbh} \right)^{-1/2}.
\label{eq:time_kozai_corrected}
\end{equation}
The period $T_\mathrm{K}$ from eqs.\
(\ref{eq:time_kozai})--(\ref{eq:time_kozai_corrected}) fits well with
results of the numerical integration shown in Figure~\ref{fig:T_K},
where we evaluate the Kozai cycle period as a function of the mass
$M_\mathrm{c}$ of the spherical cluster. $T_\mathrm{K}$ in
terms of orbital period $P$ is determined numerically by integrating the
trajectories that reach the tidal radius $\rt$ at the maximum of  the
eccentricity oscillations, i.e.\ $T_\mathrm{K}$ in Fig.~\ref{fig:T_K} is
given by the loss cone boundary trajectories. The dependency indicates
that the shape of the loss cone is also modified, and we look to this
more in the next section.

\subsection{Loss cone in the presence of Kozai's process}
\label{sec:effective_lc}
The elongated trajectories bring stars of the nuclear cluster
close to the centre. It is then natural to assume that passages near the
black hole are moments important for the evolution of the system --
feeding the black hole and triggering the accretion activity, as is
commonly described in terms of loss-cone processes (e.g.\ Frank
\& Rees \cite{frank76}; Magorrian \& Tremaine \cite{magorrian99};
Merritt \& Wang \cite{merritt05}).

A `classical' loss-cone is determined by the process of tidal
disruption of stars near the central black hole.
Low angular momentum trajectories are relevant because stars following
these orbits have their periapses within the black hole tidal radius,
\begin{equation}
\rt=\left(\frac{\mbh}{\mast}\right)^{1/3} \!\rast
\simeq 10^{3} M_4^{-2/3}
\left(\frac{\mast}{\msun}\right)^{-1/3}
\left(\frac{\rast}{\rsun}\right) \,\rg,
\label{eq:Rt}
\end{equation}
where $\mast$ and $\rast$ are the stellar mass and radius. The loss-cone
is emptied on a time-scale of $\simeq{P}$ provided the perturbing
processes are not strong enough to deflect stars from their orbits.

The above-described situation becomes more complicated when the timescales 
are modified in the presence of Kozai's mechanism. An `effective'
loss cone can be now associated with the new period $T_\mathrm{K}>P(a)$ 
operating in the system. The loss cone is formed by trajectories in
phase space plunging below $r=\rt$ by Kozai's oscillations. This view of
the influence of the process takes into consideration the fact that
orbital elements evolve in a systematical manner under Kozai's
mechanism. In other words, eccentricities are pumped up to large values
for extended periods of time, which makes the effect  distinct from the
stochastic nature of gravitational scattering. Furthermore, the
loss cone geometry is more complex because of the coupling that exists
between the mean eccentricity and inclination of the orbits.

\begin{figure*}
\includegraphics[width=\textwidth]{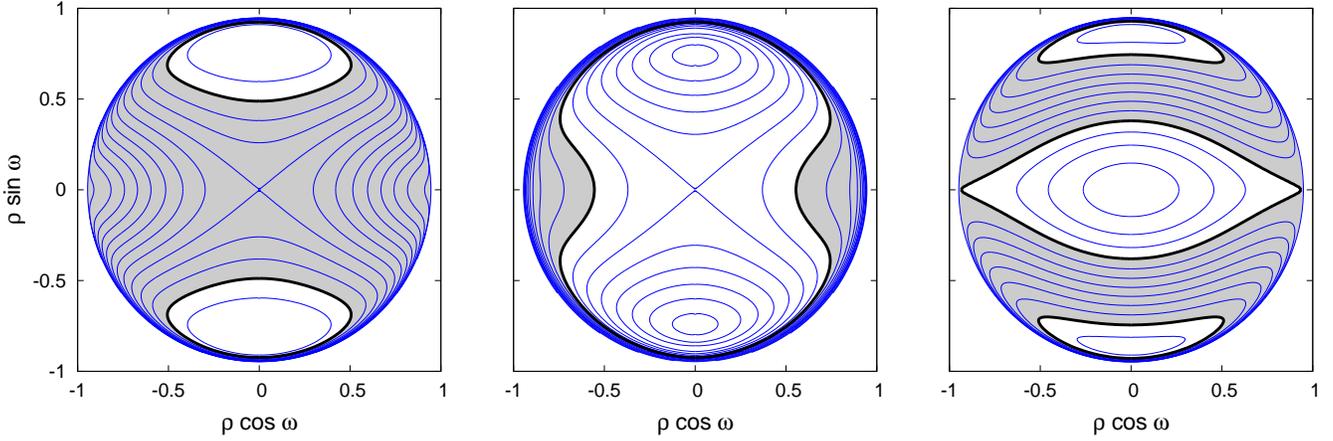}
\caption{Structure of the orbits is shown in $(\rho,\omega)$ polar plot.
To find these orbits, we assume equations of motion in the form
(\ref{eq:eqmotion}). For the gravitational field we take different terms
into account, so that the effect of perturbations becomes apparent.
Always present is the leading term of the Keplerian field (which
dominates) and the axisymmetric field $\mbf{\nabla} V_\mathrm{d}$ (which
represents the perturbation). In the left panel,  no other terms are
considered. In the middle panel, the PN1 term of the central field is
added. In the right panel, both PN1 of the centre and $\mbf{\nabla}
V_\mathrm{c}$ of the cluster are considered. High-eccentricity orbits
pass through the shaded regions. Common values of the parameters:
$M_\mathrm{c}(\rh)=\md=0.01\mbh$,  $\rd=\rh=10^5\rg$, $C_1=0.1$ and
$a=0.4\rd$.}
\label{fig:contours}
\end{figure*}

We characterize the influence of highly eccentric orbits on 
the star capture rate by the fraction of stars that
exhibit pericentre distances below a certain threshold radius,
$\rp\leq\rmin$. We denote this fraction ${\cal F}_2(\rmin;a, C_1)$ 
and write it in the form
\begin{equation}
 {\cal F}_2\equiv\frac{1}{D_2} \int_0^{\sqrt{1-C_1^2}}
 \!\!\der{e}\!\int_0^{2\pi}\!\!\der{\omega} \;\Theta(e_\mathrm{max}\! - e_\mathrm{min})
 \nonumber \; D_{\mathrm{f}}(a,C_1,e,\omega),
\label{eq:ftilda}
\end{equation}
where $\Theta$ is the Heaviside step function,
$e_\mathrm{min}\,\equiv\,1-\rmin/a$, and
$e_\mathrm{max}(a,C_1,e,\omega)$ is the maximum eccentricity reached on
a given trajectory. The normalization factor is equal to
\begin{equation}
 D_2(a, C_1) \equiv \int_0^{\sqrt{1-C_1^2}}\!\der{e}
 \int_0^{2\pi}\!\der{\omega}\; D_{\mathrm{f}}(a,C_1,e,\omega)\,.
\end{equation}
Analogically we define integrated quantities
\begin{equation}
 {\cal F}_1(\rmin;a) \equiv \frac{1}{D_1(a)}\int_0^1 \der{C_1} \;
 {\cal F}_2(\rmin;a,C_1)\,D_2(a,C_1)
\end{equation}
and
\begin{equation}
 {\cal F}(\rmin) \equiv \int_{a_\mathrm{min}}^{a_\mathrm{max}}\der{a}
 \; {\cal F}_1(\rmin;a)\,D_1(a)\,,
 \label{eq:F_tot_def} 
\end{equation}
where
\begin{equation}
 D_1(a) \equiv \int_0^1\!\der{C_1}\; D_2(a, C_1)\,.
\label{eq:D1}
\end{equation}
The function ${\cal F}(\rmin)$ can be interpreted as volume of the
loss-cone normalised to the phase space volume occupied by the stellar
cluster. (The correspondence between ${\cal F}$ and the volume of the
loss cone would be exact in the case of uniform distribution function
$D_\mathrm{f}$.)

The integrals (\ref{eq:ftilda})--(\ref{eq:D1}) can be carried out
analytically in the Keplerian case. In the non-Keplerian case, however, 
the function $e_\mathrm{max}(a,C_1,e,\omega)$ is given in an implicit form
which complicates the evaluation of the integrals. Our approach is thus
based on direct numerical integration of the equations of motion of
suitably chosen trajectories, as described below.

\subsection{Integration of the orbits}
\label{sec:num_det}
According to the averaging technique (Arnold \cite{arnold89}; Brower \&
Clemence \cite{brower61}) the mean motion is taken over orbital period
$P$. This allows us to study the long-term evolution of the osculating
parameters, $(e,\omega,a,i)$ in  an integrable (Keplerian) potential
with a small perturbation. The approach relies on the existence of a
third integral of motion (in addition to the energy and $z$-component of
the angular momentum) which represents a mean of the perturbing part of
the Hamiltonian. The orbital trajectories form a congruence of closed
curves in the space of mean eccentricity vs.\ argument of pericentre. 

We use the averaging technique to find parameters of the orbits that
reach large eccentricities. Then, for those highly eccentric orbits we
integrate the equations of motion directly. Accelerations we take in the
form
\begin{eqnarray}
\mbf{a} &\!=\!& - \frac{G\mbh}{r^3}\mbf{r} - \frac{G\mbh}{c^2\,r^3}
\left[ v^2 \mbf{r} - 4(\mbf{r{\cdot}v}) \mbf{v} - \frac{4G\mbh}{r} \mbf{r}
\right] - \mbf{\nabla} (V_\mathrm{d} + V_\mathrm{c}), \nonumber \\
& &
\label{eq:eqmotion}
\end{eqnarray}
assuming that the leading Newtonian term dominates. The bracketed term is the
PN1 correction to the gravity of the central mass (see sec.~VII of
Damour et al.\ \cite{damour91}), while $V_\mathrm{d}$ and $V_\mathrm{c}$
represent axially and spherically symmetric perturbations, respectively.

Figure~\ref{fig:contours} shows a set of tracks of different orbits with
their corresponding values of $a$ and $C_1$ kept fixed. Pericentre
argument $\omega$ stands as the polar angle, whereas 
$\rho(e)\equiv(1-\eta)^{1/2}$ is the radial coordinate; maximum 
eccentricity is then on the perimeter of these graphs,
$e_\mathrm{max}=(1-C_1^2)^{1/2}=0.995$. Two classes of orbits can be
distinguished -- those which librate around $\omega=\pi/2$ (or
$\omega=3\pi/2$) and circulating orbits for which the whole range of
$\langle0,2\pi\rangle$ is allowed. The most eccentric orbits attain 
their maximum eccentricity $e_\mathrm{max}$ at $\cos\omega=0$ in the
outer circulating region, as is evident from the plots. Hence, in order
to evaluate the fraction ${\cal F}_2(\rmin;a,C_1)$ we integrate the
orbit evolution starting from
$(e,\omega)\,\equiv\,(e_\mathrm{min},\pi/2)$. The result of integration
determines a boundary of the region occupied by trajectories with the
maximum eccentricity $\geq\,e_\mathrm{min}$. Topology of the contour
plot is crucial for the orbit behaviour. 

The shaded area shows an intersection of the loss-cone with the 
$(a,C_1)$ plane. In the example of Fig.~\ref{fig:contours} we set
$e_\mathrm{min}=0.989$ and highlight the corresponding contour. This
defines the boundary of the shaded area within which every orbit reaches
$e>e_\mathrm{min}$ at some moment of its evolution. In other words, this
area contains highly eccentric orbits and it defines the size and the
shape of the loss cone. Having the loss-cone boundary defined, we
integrate $D_\mathrm{f}(a,C_1, e, \omega)$ over the shaded region and
obtain ${\cal F}_2(\rmin;a,C_1)$.  Finally, we evaluate this function on
a grid in the $(a,C_1)$ space and integrate it numerically to find 
${\cal F}_1(\rmin;a)$ and ${\cal F}(\rmin)$.

\begin{figure*}
\includegraphics[width=\textwidth]{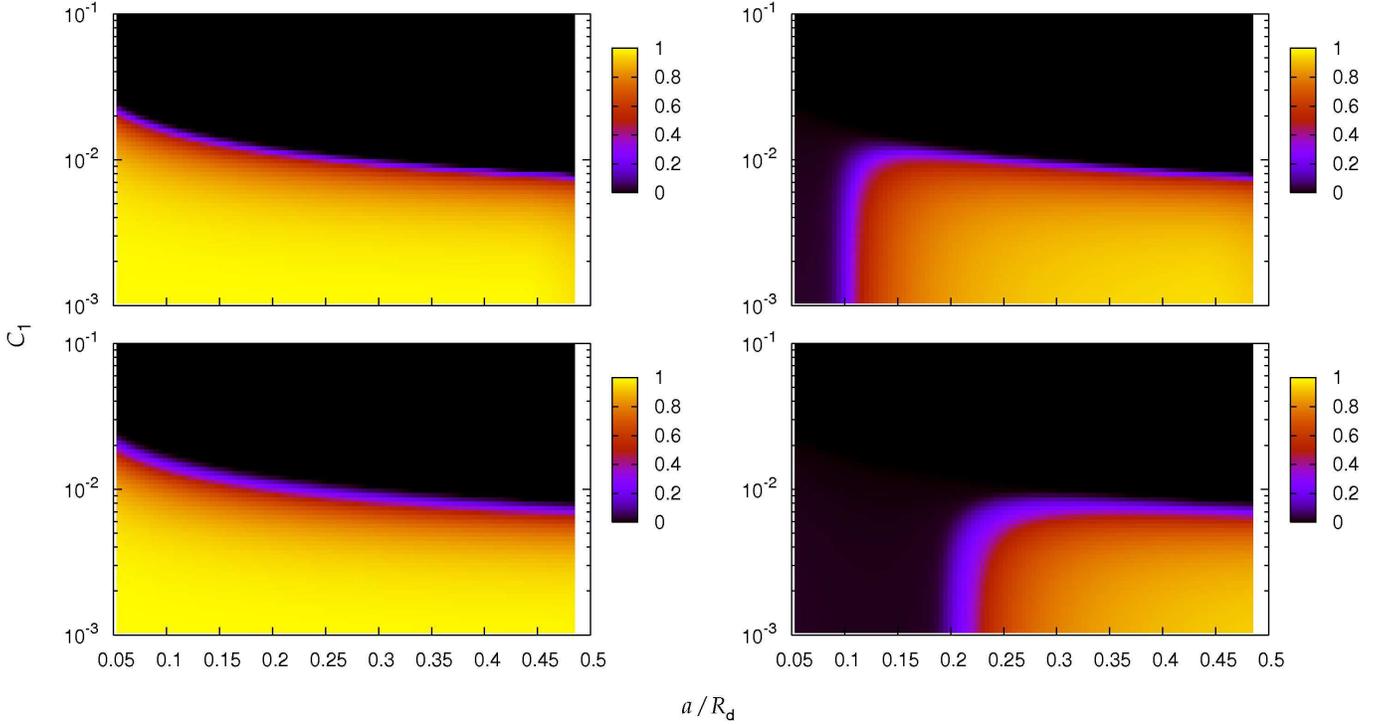}
\caption{This 
figure exhibits a clear distinction between the cases with  (on the
right) and without (on the left) the relativistic  pericentre advance
taken into account. We draw the fraction  ${\cal F}_2(\rmin;a,C_1)$ of
stars from the ensemble (\ref{eq:df}) that plunge below $\rmin=100\rg$;
$a$ and $C_1$ are constants of motion. The central black hole mass
$\mbh=3.6\;10^6\msun$, the disc mass $\md=0.1\mbh$. In the bottom panels
we assumed the central gravitational field perturbed by a narrow ring.
In the upper panels, the case of a constant-density disc was considered.
The left column is for the Newtonian potential of the central source,
while on the right side we employed the post-Newtonian (PN1) model.
Value of ${\cal F}_2$ is encoded by a colour palette, ranging from zero
to unity. Therefore, this figure demonstrates how sensitive the value of
${\cal F}_2$ is with respect to the pericentre precession and the
eccentricity oscillations -- the two effects which act against each
other. Notice a sharp drop of ${\cal F}_2$ below a certain critical
value of $a=a_\mathrm{t}$ (in the panels on the right side). This break
emerges once the pericentre advance is taken into account; it does not
occur in the left panels because these were computed in purely Newtonian
gravity.}
\label{fig:ratio_perturb}
\end{figure*}

The three panels exhibit different possible topologies of the tracks of
orbits that arise due to the perturbing forces. In order to construct
these plots, the central Keplerian potential was superposed with the
perturbation $\mbf{\nabla} V_\mathrm{d}$ (left); with PN1 correction and
$\mbf{\nabla} V_\mathrm{d}$ terms added (middle); and with PN1
correction and $\mbf{\nabla} (V_\mathrm{d} + V_\mathrm{c})$ terms taken
into account (right).

In a conservative system, which is what we consider in 
Fig.~\ref{fig:contours}, star tracks remain attached to the contours.
However, additional dissipative processes might allow adiabatic
evolution of the orbital parameters, so that stars can slide gradually
across the level surfaces on very long time-scales. This may cause a
decay of the orbits and help to bring stars to the centre. We do not
consider such effect here, but see \v{S}ubr \& Karas (\cite{subr05})
where Kozai's resonance mechanism {\em and} the hydrodynamical
(dissipative) drag are both taken into account.

\section{Results}
As mentioned above, Kozai's resonance mechanism is active when the
Newtonian central potential is perturbed by an axisymmetric term -- i.e.
the disc potential in our case, however, the effect is known to be
fragile with respect to various other perturbations that may influence
the motion of stars. Namely, it is suppressed when the central field is
non-Newtonian,  leading to the precession of the orbits. Therefore we
want to clarify whether Kozai's mechanism is still relevant for the
long-term dynamics  of nuclear stars even if relativistic pericentre
advance and the effect  of the nuclear cluster are taken into account.

\subsection{Fractional probabilities}
First we neglect all perturbing terms in the gravitational field and
assume purely Keplerian motion. The maximum
eccentricity along each trajectory is simply
$e_\mathrm{max}(a,C_1,e,\omega)\,\equiv\,e=\mbox{const}$.
Assuming the distribution (\ref{eq:df}), fractional probabilities
${\cal F}_i$ can be then written in the explicit form:
\begin{equation}
{\cal F}_2(\rmin;a,C_1)=\frac{\eta_\mathrm{min}-C_1}{1-C_1}\,
\Theta(\eta_\mathrm{min} - C_1)\,,
\label{eq:Fac_kepler}
\end{equation}
where $\eta_\mathrm{min}\equiv\,(1-e_\mathrm{min}^2)^{1/2}$,
\begin{equation}
{\cal F}_1(\rmin;a)=\eta^2_\mathrm{min} = \frac{2\rmin}{a} - \frac{\rmin^2}{a^2}\,,
\label{eq:Fa_kepler}
\end{equation}
\begin{eqnarray}
{\cal F}(\rmin) &=& \frac{10\,\rmin\left(a_\mathrm{max}^{1/4}
 - a_\mathrm{min}^{1/4}\right) + \frac{5}{3} \rmin^2\left(a_\mathrm{max}^{-3/4}
 - a_\mathrm{min}^{-3/4}\right)}{a_\mathrm{max}^{5/4} - a_\mathrm{min}^{5/4}} 
 \label{eq:F_tot_kepler1}
\\
 &\simeq& 10\, \rmin \, a_\mathrm{max}^{-1}\,.
 \label{eq:F_tot_kepler}
\end{eqnarray}
To derive the approximation (\ref{eq:F_tot_kepler}) we assumed 
$\rmin\ll a_\mathrm{min} \ll a_\mathrm{max}$.

Under an axially symmetric perturbation of the central potential the
total angular momentum of orbits is no longer conserved, nevertheless,
its $z$-component remains an integral of motion. This results in
oscillations of eccentricity. An upper estimate of the corresponding
probabilities can be derived from the loss-cone condition,
$L_z<L_\mathrm{max}$, which can be expressed as
$e_\mathrm{max}(a,C_1,e,\omega) = (1-C_1^2)^{-1/2}$. Then,
\begin{equation}
{\cal F}_2(\rmin;a,C_1) = \Theta(\eta_\mathrm{min} - C_1)\,,
\label{eq:Fac_kozai}
\end{equation}
\begin{equation}
{\cal F}_1(\rmin;a) = 2\eta_\mathrm{min} - \eta_\mathrm{min}^2 \simeq
2\left(2\rmin \, a^{-1}\right)^{1/2}\,,
\label{eq:Fa_kozai}
\end{equation}
and
\begin{equation}
 {\cal F}(\rmin) \simeq \frac{10}{3} \frac{\left(2\rmin\right)^{1/2}
 \left(a_\mathrm{max}^{3/4} - a_\mathrm{min}^{3/4}\right)}{a_\mathrm{max}^{5/4} -
 a_\mathrm{min}^{5/4}}
\label{eq:F_tot_kozai1}
 \simeq 4.71 \left(\frac{\rmin}{a_\mathrm{max}}\right)^{1/2}\!\!.
\end{equation}
The last term on the right-hand side gives the values two to three
orders of magnitude larger than those which follow from
eq.~(\ref{eq:F_tot_kepler}). The estimate 
(\ref{eq:F_tot_kozai1}) is good provided the Kozai mechanism is the only
one determining the orbital changes. However, this resonance
mechanism may be diminished by other perturbations, so the actual value
of ${\cal F}(\rmin)$ should be somewhere in between the two estimates.
We will now discuss the importance of damping effects.

\begin{figure*}
\includegraphics[width=\textwidth]{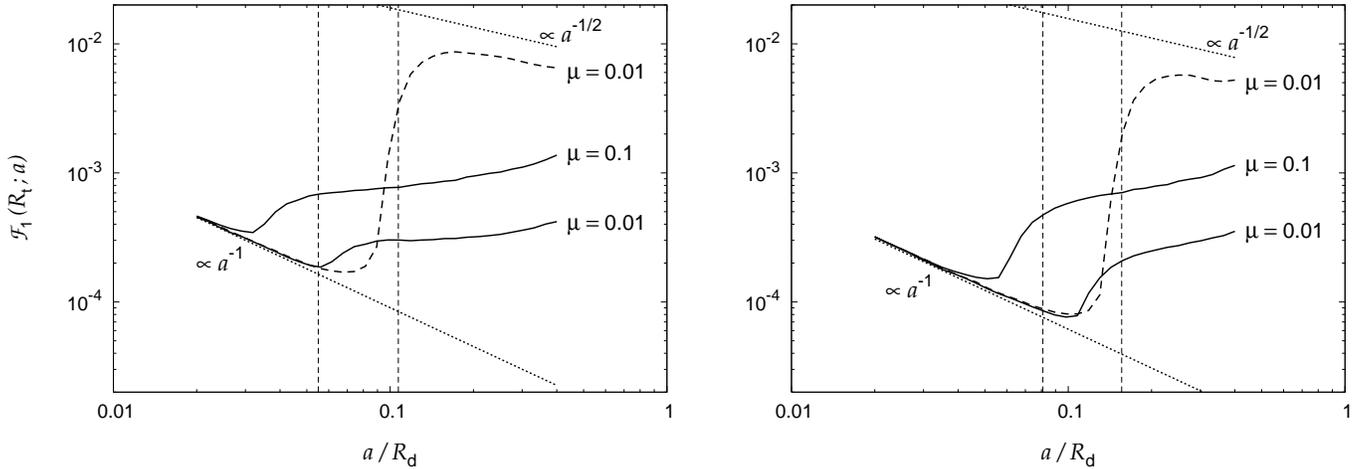}
\caption{Fraction 
${\cal F}_1(\rt;a)$ of stars plunging below the black hole  tidal
radius. Stellar orbits are perturbed by a massive ring of radius $\rd$ 
and mass $\md=\mu\mbh$ (indicated with curves). The dashed line is for
the PN1 approximation of the central mass gravity field; the effect of
the outer stellar cluster is neglected. Solid lines represent two cases
in which the cluster is taken into account (they differ by the disc mass
parameter $\mu$). Dotted are two power-law lines corresponding to the
lower and the upper analytical estimates, eqs.~(\ref{eq:Fa_kepler}) and
(\ref{eq:Fa_kozai}), respectively. Vertical dashed lines indicate the
values of $a_\mathrm{t}$ (given by eq.~(\ref{eq:a_term}) with $\mu=0.01$
and $0.1$, respectively); these are critical semiaxis values below which
Kozai's phenomenon is suppressed. For $a>a_\mathrm{t}$ the eccentricity
oscillations persist. Parameter values assumed in this plot are: in the
left panel $\mbh=10^4\msun$; in the right panel $\mbh=10^5\msun$.}
\label{fig:Fa}
\end{figure*}
The relativistic advance of pericentre has an overall tendency of
decreasing the maximum eccentricity over the Kozai cycle. Below a
certain threshold value of $a$ the maximum eccentricity is not
sufficient to bring stars inside the tidal radius; the loss-cone 
condition is then $L<L_\mathrm{max}$. We estimate this terminal value as
\begin{equation}
 a_\mathrm{t}^7\simeq{\textstyle{\frac{32}{9}}}\,\rd^6\,\rg^2\,\rmin^{-1}
 \,\mu^{-2}
\label{eq:a_term}
\end{equation}
(see Appendix~\ref{sec:a_t}). For $a>a_\mathrm{t}$ the maximum
eccentricity allowed by the relativistic corrections is large enough, so
that the pericentre gets below $R_\mathrm{min}$. Hence, the fraction of
tidally disrupted stars is determined solely by the
estimate~(\ref{eq:Fa_kepler}) for $a<a_\mathrm{t}$, while above
$a_\mathrm{t}$ the fraction increases abruptly and saturates almost at
the value (\ref{eq:Fa_kozai}).

\subsection{Stars plunging below a threshold radius}
In figure~\ref{fig:ratio_perturb} we plot the fractional probability
${\cal F}_2(\rmin;a,C_1)$, as we obtained it numerically. Comparison
between different panels clearly demonstrates the reason for the concern
that the pericentre advance of stellar orbits might completely erase the
effect of resonance.

One can see that Kozai's mechanism indeed sets stars on eccentric
trajectories reaching small radii and that the efficiency of the process
increases with $a$ decreasing down to a certain limiting value. We
considered both the Newtonian and the post-Newtonian models of the
central field, so the effect of the relativistic pericentre advance can
be distinguished. We find that ${\cal F}_2$ is raised by a factor of
$\simeq100$ due to the Kozai mechanism and it drops at small semi-axis; 
the exact value of $a$ where that break occurs depends on the adopted
form of the perturbing potential. Comparison of the panels reveals that
the disc-like source of the gravity competes more successfully with the
relativistic effect than a narrow ring of the same mass.

The mean potential of the extended cluster of stars also causes secular
precession of the orbits but this time it alters the results in a
different way than the relativistic pericentre advance by attenuating
the Kozai oscillations in the whole range of semi-major axes. 
Nevertheless, it still allows a substantial fraction of orbits to reach
high eccentricities.

The fraction ${\cal F}_1(\rt;a)$ is shown in figure~\ref{fig:Fa}. We
find the damping effects to be weakened if the mass of the axisymmetric
component is higher, as expected. By increasing the mass parameter $\mu$
the curve of ${\cal F}_1(\rt;a)$ gets gradually shifted to larger
values. Simultaneously, $a_\mathrm{t}$ is diminished. As consequence of
these dependencies, more stars reach the tidal radius for lower mass of
the central body than for higher masses. For the sake of definiteness,
we assume solar-type stars ($\mast=1\msun$ and $\rast=1\rsun$) in the
evaluation of $\rt$ in eq.~(\ref{eq:Rt}).

\begin{figure*}
\includegraphics[width=\textwidth]{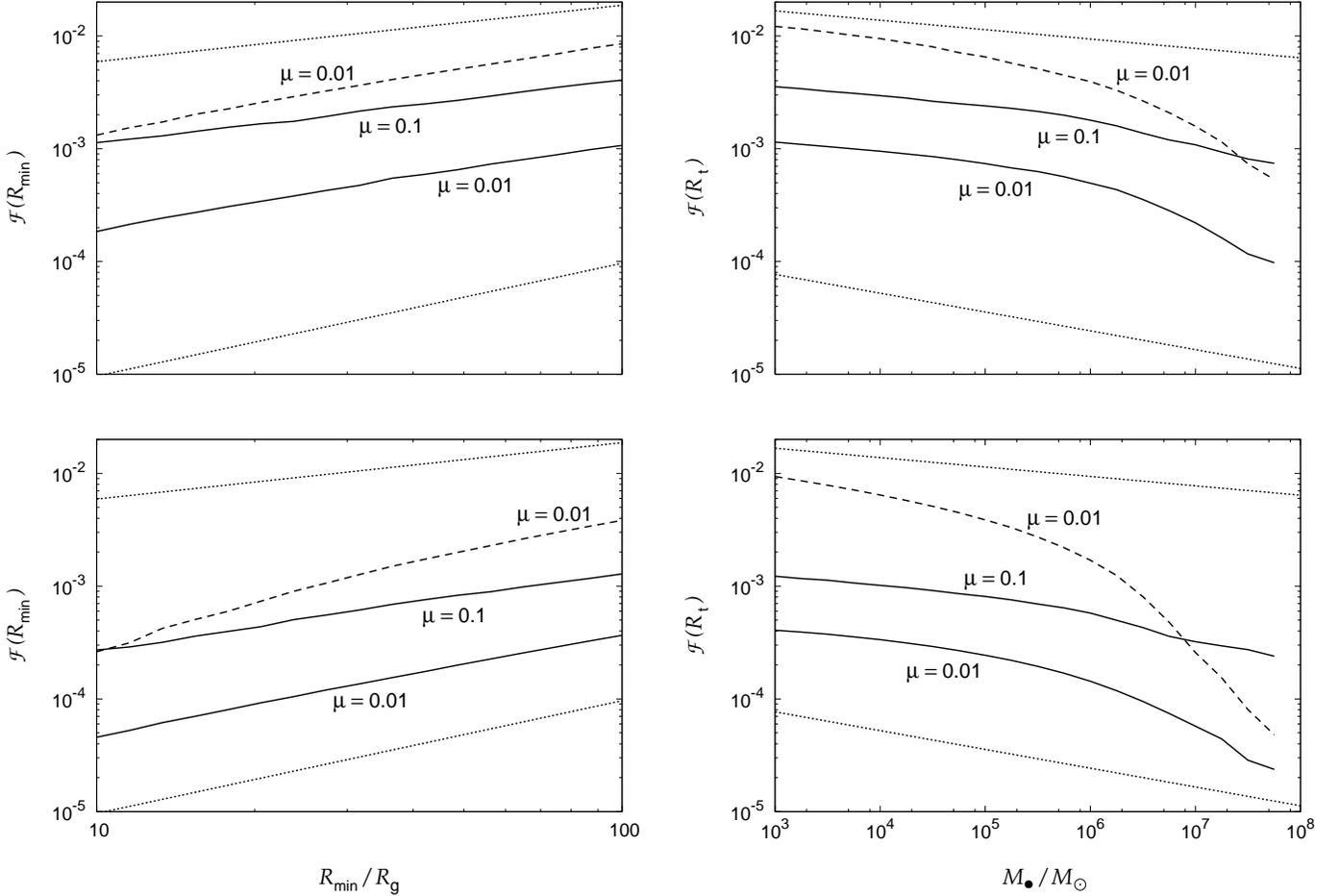}
\caption{Fraction 
${\cal F}(\rmin)$ of stars reaching the pericentre below a given radius
$r=\rmin$. We assumed perturbing potentials $V_\mathrm{d}(R,z)$ of the
disc with constant surface density (top panels) and a narrow ring
(bottom panels). Left: the functional dependence of probability ${\cal
F}(\rmin)$ (solid lines).  Parameters compatible with Sgr~A$^\star$ were
assumed: $\mbh=3.5\times10^6\msun$, $0.01\leq\mu\leq0.1$,
$\rh\simeq\rd=1.2\times10^7\rg=2$~pc, and $0.04\rh\leq{a}\leq 0.4\rh$.
Right: the mass dependence, ${\cal F}(\rt(\mbh))$, computed according to
eqs.~(\ref{eq:Rh}) and (\ref{eq:Rt}). In all panels dotted lines
correspond to the analytical estimates (\ref{eq:F_tot_kepler1}) and
(\ref{eq:F_tot_kozai1}); values given by the latter correspond within a
factor of $\lesssim 2$ to the case omitting both the relativistic
pericentre advance and the stellar cluster reducing of the Kozai
oscillations. Dashed lines represent models with only the relativistic
precession switched on ($\mu=0.01$), while the solid lines include also
the precession from the cluster potential (two cases are shown,
$\mu=0.01$ and $0.1$).}
\label{fig:F}
\end{figure*}

Finally, the overall fraction ${\cal{}F}(\rmin)$ is shown in
figure~\ref{fig:F}. The limits on semi-major axes range are set mainly
by numerical arguments: the lower limit results from a sharp increase of
the Kozai time $T_\mathrm{K}$ when $a$ is decreased. Below
$a_\mathrm{min}=0.04\rh$ the orbits would contribute to the tidal
disruption rate only marginally. The upper limit has to be set due to
limitations of our approach -- at $r\simeq\rd$ the orbits exhibit
chaotic motion and they no more follow the curves of
figure~\ref{fig:contours}. We assume that these orbits will lead to the
disruptions and, therefore, results presented here should be considered
as {\em lower estimates}.

Line types help us to distinguish the importance of different effects in
this plot. In particular:
\begin{list}{}{%
\setlength{\leftmargin}{2em}
\setlength{\labelwidth}{2em}} 
\item[{\makebox[1.5em][r]{(i)}}]Kozai's oscillations rise the value
of $\cal F$ with respect to an unperturbed ($\mu=0$) case;
\item[{\makebox[1.5em][r]{(ii)}}]PN1 corrections to the central
(Keplerian) potential reduce $\cal F$ by factor $\simeq2\div5$
(the expected rate of tidal disruptions is diminished accordingly); 
\item[{\makebox[1.5em][r]{(iii)}}]precession due to self-gravity
of the cluster decreases $\cal F$ further down by another factor
$\simeq10$. 
\end{list}

As in the previous figures, it is clearly visible that the potential of
the extended source, i.e.\ the disc, competes more successfully against
damping than the ring. Increasing the mass of the axial perturbation not
only rises $\cal F$ to higher values, but it also changes the slope.
That tendency can be attributed to larger terminal value $a_\mathrm{t}$
of the semi-major axis, above which the relativistic damping is
negligible.

The effect of relativistic pericentre advance increases with the mass of
the black hole. The right panels of Fig.~\ref{fig:F} demonstrate that
${\cal{}F}$ is anti-correlated with $\mbh$ and grows towards less massive
black holes (i.e.\ when going from SMBH to IMBH).

\section{Discussion}
We examined the idea of Kozai's effect assisted by an accretion disc as
the mechanism enhancing the disruption rate of stars by a central black
hole. This process is more efficient for intermediate-mass black holes
than for supermassive ones. In our calculations we used PN1
approximation of the central field. We performed these calculations also
within the pseudo-Newtonian (Paczy\'nski--Wiita \cite{paczynski80})
framework (as a check of the orbits very near horizon) with very similar
conclusions.

The results for ${\cal F}(\rt)$ can be directly interpreted as fraction
of stars populating the loss-cone when an axisymmetric perturbation is
applied on the cluster. We suggest that this picture is relevant in a
system where gas clouds form an embedded disc-like structure with a
non-negligible total mass near the sphere of dominance of a central
black hole. 

\subsection{Disturbing processes}
The view presented here is valid as far as other processes,
acting on time scales shorter than $T_\mathrm{K}$, do not manage to
expell the stars from the effective loss cone. This would inhibit the
Kozai process. One of the generic processes that may interfere with the
influence of the axisymmetric perturbation is the two-body relaxation
within the stellar cluster.

Gravitational relaxation brings the diffusion time-scale into the problem,
$T_\theta\,\equiv\,\theta_\mathrm{lc}^2 T_\mathrm{r}$, where
\begin{equation}
\theta_\mathrm{lc} = \sqrt{\rt/r} \simeq \sqrt{{\cal F}(\rt)} \ll1
\end{equation}
is an angular extent of the loss cone and
\begin{equation}
T_\mathrm{r}= \frac{\sigma^3}{G^2C\ln\Lambda\,M_{\ast}^2n_{\ast}}\,.
\end{equation}
Here, $n_{\ast}$ is the number density of the stellar system,
$\ln\Lambda$ is Coulomb logarithm and $C$ is a constant (in usual
notation, $C\ln\Lambda\sim10^2$; Spitzer \cite{S87}). In a cusp
described by the Bahcall--Wolf distribution, velocity dispersion is
comparable with Keplerian velocity and the number density of stars
decreases with radius as $n_\ast(r)\propto r^{-7/4}$. Characterizing the
number density by the stellar mass $M_\mathrm{c}$ enclosed  within the
radius $\rh$, we get
\begin{equation}
 T_\mathrm{r} \simeq 10^{6}\, M_4^2\,
 \left(\frac{M_\mathrm{c}}{\mbh}\right)^{-1}
 \left( \frac{M_\ast}{M_\odot}\right)^{-2}
 \left( \frac{\rh}{10^8\rg}\right)^{5/4}
 \left(\frac{r}{10^8\rg} \right)^{1/4} {\rm yr}\;.
\label{eq:time_relax_bw}
\end{equation}

The effective loss cone which we have invoked in
Sec.~\ref{sec:effective_lc} generally exceeds $\theta_\mathrm{lc}$. From
Fig.~\ref{fig:F} we see that its volume  ${\cal F}(\rt)$ is (for
$\mu=0.1$ and a ring source of the perturbation) by factor $>10$
larger than it is in an unperturbed system. Hence, appropriate
$\theta_\mathrm{lc}^\prime$ must be by a factor $\gtrsim3$ larger than
$\theta_\mathrm{lc}$. However, the geometry of the effective loss cone
is more complicated. In particular, it is narrower at $\omega=\pi/2$
where it actually coincides with the boundary of the classical loss
cone. Therefore, at $\omega$ corresponding to the minimum eccentricity
the new loss cone must be sufficiently wider so that its volume can
exceed by a factor of ten the classical one. We estimate 
$\theta_{\mathrm{lc}}^{\prime} \simeq 10 \theta_\mathrm{lc}$, which  is
also in accordance with our experience that the pericentre value on the
boundary orbit of the effective loss cone typically changes by more than
two orders of magnitude. The time scale on which stars diffuse across
the  effective loss-cone is then
\begin{equation}
 T_\theta^\prime \simeq 100\, T_\theta \simeq 0.04\, M_4^{3/2}
 \left( \frac{M_\mathrm{c}}{\mbh} \right)^{-1}
 \left( \frac{M_\ast}{M_\odot} \right)^{-1}
 \left( \frac{r}{\rh} \right)^{-9/4} P.
\end{equation}
By comparing $T_\mathrm{K}$ and $T_\theta^\prime$ we deduce that for
$M_4\gtrsim10$ the Kozai mechanism affects the angular momentum on a
significantly shorter time-scale and we may consider the effective loss
cone to be emptied by this process.  

For lower BH masses the interplay
of Kozai mechanism and the two-body relaxation is more complicated,
however, we still expect that {\em systematic\/} character of the
angular momentum changes due to the Kozai mechanism will {\em enhance\/}
the rate of tidal disruptions. A definitive study of these systems is
probably not possible without employing high precision $N$-body
integrators.

We imagine that consumption of stars by the central IMBHs is a transient
process which enhances their activity during those periods when stars
are supplied into the loss-cone and then efficiently brought onto
eccentric orbits. Permanent replenishment of this larger loss cone is
not critical for the process considered here. Nevertheless, we may
assume that the larger loss cone is replenished more frequently (also by
other processes than the steady state relaxation -- for example  the
effect of Brownian motion of the central black hole, or a secondary
black hole can contribute). However, details of the loss-cone refilling
are beyond the scope of our present paper. It will be interesting to
compare these speculations with the results of $N$-body simulations that
have been recently applied to explore the tidal processes and their
effect on IMBHs formation and feeding in dense star clusters (see
Baumgardt et al.\ \cite{baumgardt06}).

To achieve a more complete description one has to invoke other diffusive 
processes acting together with the gravitational relaxation (e.g.\ the
hydrodynamical drag by the disc gas that continuously helps re-filling
the loss-cone; see e.g. Karas \& \v{S}ubr \cite{karas01} for relevant
time-scales) but, again, this is beyond the scope of our paper.

\subsection{Dependence on the black hole mass}
Let us consider two types of object which differ from each other by
the black hole mass -- (i)~a supermassive black hole in the Galaxy
center, and (ii)~hypothetical intermediate-mass black holes that might
reside in cores of dense star clusters.

Precise tracking of the proper motion of individual stars is currently
possible within an arcsecond area around the Galaxy centre (Genzel et
al.\ \cite{genzel03}; Ghez et al.\ \cite{ghez03}), and so the idea of
applying our calculation to Sgr~A$^\star$ is quite natural and it raises
a question about the origin of various perturbations that may act on
stellar motion. Currently, $\md\simeq0.1\mbh$ can be set as an upper
estimate of the mass of the molecular circumnuclear disc which extends
from $\sim1.5$~pc to $3\div4$~pc (Christopher et al.\
\cite{christopher05}), i.e.\ on the outer edge of the
black hole's sphere of influence. This is compatible with a narrow ring
of radius $\rd=\rh\simeq 2\mathrm{pc}$, and so we can directly evaluate
the expected number of tidal disruptions. From Fig.~\ref{fig:F} we read
that the fraction $\simeq 5\times10^{-4}$ of the total number of stars 
in the cluster with $a_\mathrm{max}<0.4\rd$ exhibit sufficiently large
eccentricity oscillations. This corresponds to the mass in stars about 
$\simeq 0.1\mbh \simeq 3\times10^5
\msun$, thence we obtain $\simeq150$ tidal disruption events per
$T_\mathrm{K} \simeq 10\;\mathrm{Myr}$.

So far we assumed that the source of the gravitational perturbation was
a gaseous disc or a torus, however, its origin could be different. In
the Sgr~A$^\star$ there is a disc of young stars (Levin \& Beloborodov
\cite{levin03}; Paumard et al.~\cite{paumard06}) and, possibly, a
remnant gas from which these stars had been born. To model their effect
we consider a thin disc of mass $0.01\mbh$ and surface density $\propto
R^{-2}$ extending between  $0.03\div0.3\;\mathrm{pc}$. We found that
$\simeq 2$\% of stars from the region $<0.3\mathrm{pc}$ can undergo the
oscillations that bring them down to $\rt$. This means that during the
period of $T_\mathrm{K}\lesssim 1\mathrm{Myr}$ after the formation of
the gaseous/stellar disc there may have occurred up to $\simeq100$ more
tidally disrupted stars compared to the result of calculations
neglecting the effect. The phase of enhanced disruption rate may become
prolonged if the orientation of the disc varies in time. This could be
caused by precession in the outer galactic potential.

In the case of IMBHs the resonance mechanism complements the role of
stellar encounters in the process of feeding the black hole. If the
total mass of the cluster is about $\mbh\simeq10^4\msun$, we
expect the enhancement of the disruption rate $\simeq1\msun$ per
$10^3$~yrs (with $\mu=0.1$). Disrupted stars then provide material for
accretion, thereby triggering a transient luminous phase of the object. 

We remark that the model has only formal validity near the lower end of
the black-hole mass range because in this case the accretion rate of
$\simeq10^{-3}\mbh\,\mbox{yr}^{-1}$ is comparable with the mass carried by
individual stars; other processes must be important under such
circumstances (which is consistent with the expectation that the
accretion of stellar material onto IMBH is a non-steady process).
Therefore, the scenario outlined above conforms to the assumption that
ULXs are a transient phenomenon for which accretion of the surrounding
medium is essential.
 
\section{Conclusions}
Disruption of stellar bodies and subsequent accretion of the remnant gas
are among likely mechanisms feeding black holes that are embedded in a
dense cluster. We discussed one of the channels that may contribute to
this process. Kozai's mechanism can enhance the rate of such events,
trigger the episodic gas supply onto the black hole, and, consequently,
strengthen the activity of the system by raising the accretion rate.
The process acts at characteristic time-scale of the Kozai cycle --
typically $T_\mathrm{K}\simeq10^5$~yrs for IMBHs, during which the loss
cone is depleted. 

Stars on highly elongated orbits are susceptible to tidal disruption and
hence they provide a natural source of  material to replenish the inner
disc. We have demonstrated that this phenomenon operates in the system
even if it is disturbed by the relativistic  pericentre precession and
gravity of the nuclear cluster. Therefore we can conclude that the
presence of a gaseous disc of a small but non-zero mass, $0<\md<\mbh$,
helps dragging stars to the black hole, thereby feeding the centre and
simultaneously providing material that sustains and replenishes the disc
itself. 

\begin{acknowledgements}
We thank Marc Freitag for helpful discussions about the gravitational 
influence of the cluster and Holger Baumgardt about the possible role of
IMBHs in ULXs. This work was supported by the Centre for Theoretical
Astrophysics in Prague (ref.\ LC06014) and the DFG Priority Program 1177 
`Witnesses of Cosmic History: Formation and Evolution of Black Holes, 
Galaxies and Their Environment'. The Astronomical Institute of the Academy 
of Sciences is  financed via Ministry of Education project ref.\ 
AV0Z10030501. VK gratefully acknowledges the continued support from
the Czech Science Foundation (ref.\ 205/07/0052).
\end{acknowledgements}

\appendix
\section{The role of GR pericentre advance}
\label{sec:a_t}
Here we want estimate the region of the parameter space where the
general relativistic pericentre advance dominates over the Kozai effect.
The quadrupole approximation leads to evolutionary equations (Kiseleva
et al.~\cite{kiseleva98}; Blaes et al.~\cite{blaes02}):
\begin{eqnarray}
T_\mathrm{K}\,\eta\,\,\frac{\der i}{\der t} &\bs=\bs&
 -\frac{15}{8}\,e^2\,\sin2\omega\,\sin i\,\cos i\,,
 \label{eq:didt} \\
T_\mathrm{K}\,\eta\,\,\frac{\der e}{\der t} &\bs=\bs&
 {\frac{15}{8}}\,e\,\eta^2\,\sin2\omega\,\sin^{2}i\,,
 \label{eq:dedt} \\
T_\mathrm{K}\,\eta\,\,\frac{\der\omega}{\der t} &\bs=\bs&
 \frac{3}{4}\left\{ 2\eta^2+5\sin^{2}\omega\left[e^{2}-\sin^{2}i\right]\right\}
 + T_\mathrm{K}\, \frac{\rg}{a\,\eta}\,\frac{6\pi}{P}\,.
\label{eq:doKdt}
\end{eqnarray}
The last term on the right-hand side of eq.~(\ref{eq:doKdt}) includes
the relativistic pericentre advance.

Equations (\ref{eq:didt})--(\ref{eq:doKdt}) imply conservation of
\begin{equation}
C_1 =\eta\,\cos i 
\end{equation}
and
\begin{equation}
C_2 =\left(5\sin^2 i\, \sin^2\omega - 2\right)\,e^2
 -\frac{8\,\rd^3\,\rg}{\mu\, \eta\, a^4}\,.
\end{equation}
Solutions of two types can be found -- circulating vs.\ librating --
depending on the motion constants $a$, $C_1$, and $C_2$. The circulation
region exists always while the libration region occurs typically for
small values of $C_1$. The two categories are separated by the
separatrix curve that passes through $e=0$ point in the plane of
$(e,\omega)$ polar coordinates. This can be used to determine the value
of $C_2$ on the separatrix:
\begin{equation}
C_{2,\mathrm{s}}= -\frac{8\rd^3\,\rg}{\mu\,a^4}\,.
\label{eq_C2_s}
\end{equation}
Maximum eccentricity on the separatrix is given by equation
\begin{equation}
5\left(1-\frac{C_1^2}{\eta_\mathrm{s}^2}\right)
-\frac{8\rd^3\,\rg}{\mu\,a^4}\frac{1}{\eta_\mathrm{s}(1+\eta_\mathrm{s})}
= 2\,,
\end{equation}
which for $\eta_\mathrm{s}\ll 1$ simplifies to the form 
\begin{equation}
3\eta_\mathrm{s}^2 - \frac{8\rd^3\,\rg}{\mu\, a^4}\,\eta_\mathrm{s}
= 5C_1^2.
\label{eq:eta_s}
\end{equation}

We are interested in $C_1$ small. For $C_1 \lesssim0.01$, the circulating
solutions cover about half of the $(e,\omega)$ parameter space. These
orbits acquire eccentricities larger than $e_\mathrm{s}$ at some moment
during the orbit evolution. Hence, when $e_\mathrm{s}(a,C_1)>1-a/\rmin$
the fraction ${\cal F}_2(\rmin;a,C_1)$ is close to the corresponding
estimate (\ref{eq:Fac_kozai}) which assumes that {\em all\/} orbits with
given values of $a$ and $C_1$ reach the pericentre below $\rmin$.
Setting $\eta_\mathrm{s}^2=1-e_\mathrm{min}^2\simeq2\rmin/a$,
eq.~(\ref{eq:eta_s}) gives an implicit formula for the boundary of these 
two regions in ($a,C_1$) plane. We can solve (\ref{eq:eta_s}) for
$C_1=0$ to determine the terminal value (\ref{eq:a_term}) of semi-major
axis $a_\mathrm{t}$. Below this value the relativistic pericentre
advance inhibits the Kozai mechanism.

\end{document}